\documentclass[12pt, letterpaper]{article}

\usepackage{amsmath,amssymb}
\usepackage{cite}
\usepackage{fancyhdr}
\usepackage[top=1in, bottom=1in, left=1in, right=1in]{geometry}
\usepackage{graphicx}
\usepackage{hyperref}

\numberwithin{equation}{section}
\date{}

\begin{document}
\title{{\rm\footnotesize \qquad \qquad \qquad \qquad \qquad \ \qquad \qquad \qquad \ \ \ \ \ \                      RUNHETC-2024-15
}\vskip.5in    Dumbbell Fermions and Fermi-Pauli Duality}
\author{Tom Banks\\
NHETC and Department of Physics \\
Rutgers University, Piscataway, NJ 08854-8019\\
E-mail: \href{mailto:tibanks@ucsc.edu}{tibanks@ucsc.edu}
\\
\\
}

\maketitle
\thispagestyle{fancy} 

\begin{abstract}  We use the Kantor-Susskind\cite{kantsuss} model of fermions as "dumbbells" connecting points on a cubic lattice to points on its dual, to define a duality between local fermionic models invariant under a $Z_2$ gauge symmetry and models of bosonic variables (generalizations of Pauli matrices) defined on the lattice.
\normalsize \noindent  \end{abstract}


\newpage
\tableofcontents
\vspace{1cm}

\vfill\eject
\section{Introduction}

There have been many attempts to generalize bosonization beyond the $1 + 1$ dimensional context in which it was first discovered.  In continuum field theory, these usually involve the idea of gauging the $Z_2$ symmetry $( - 1)^F$ and this program has succeeded for a variety of models with Chern-Simons gauge fields in $2 + 1$ dimensions\cite{cs}.  We also know how to construct fermions from bosons using the Dirac quantization condition\cite{hasenfratz} in $3 + 1$ dimensions.   

Attempts to find an analog of the Jordan-Wigner transformation in lattice models have been somewhat less successful\cite{latticejw}.  In this note we want to report on a simple general procedure, which seems to do the trick.  Our basic starting point is the observation\cite{tbfermipauli1} that Hamiltonian lattice gauge theory for the discrete groups $Z_N$ is most elegantly formulated by assigning the canonical conjugate $V_{S(l)}$ of $U(l)$ to the $d - 1$ dimensional face $S(l)$ of the hypercube on the dual lattice, that is pierced by the lattice link $l$.  We have
\begin{equation} U(l) V_{S(l)}  = V_{S(l)} U(l) e^{2\pi i /N} . \end{equation}
We will be interested only in the case $N = 2$.  We take our lattice to be toroidal, as is the dual lattice.  The requisite identifications are shown in two dimensions in Figure 1.  
\begin{figure}[h]
\begin{center}
\includegraphics[width=01\linewidth]{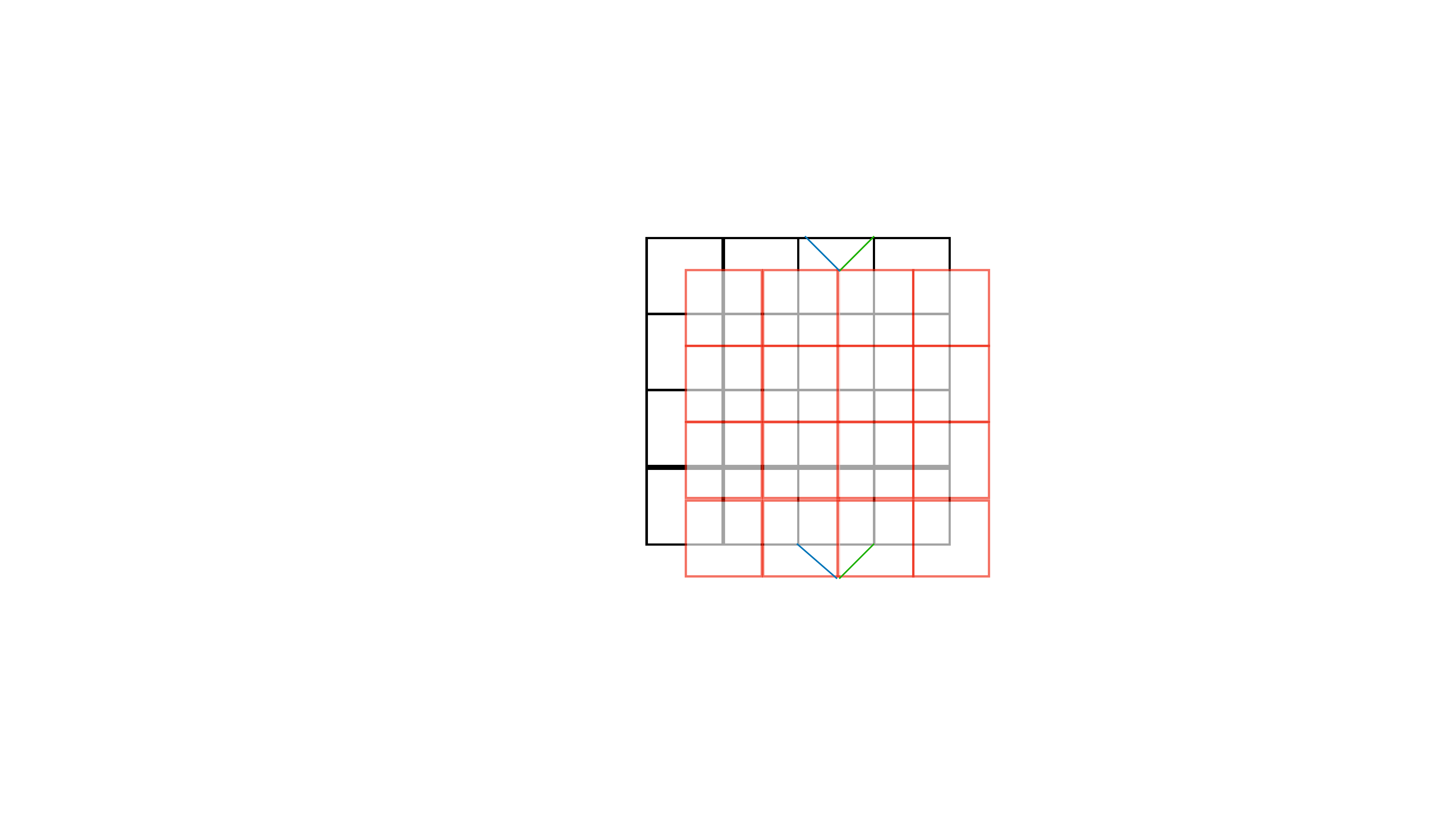}

\caption{Periodic boundary conditions: the Clifford variables at the two ends of each pair of a line and a dual line are identical.} 
\label{fig:fermipaulifig1}
\end{center}
\end{figure}

We can also denote a link variable $U(l)$ by the pair of lattice points $U(m,n)$ that it connects.  Then the $Z_2$ gauge transformation is
\begin{equation} U(m,n) \rightarrow W(m) U(m,n) W(n)\end{equation} where $W(n)$ belongs to an independent $Z_2$ group at each point.  There is also a higher form gauge symmetry associated with the $d - 1$ dimensional hypercube variables $V(S(l))$.  This is only a symmetry if we restrict attention to the topological $Z_2$ gauge theory\cite{kapseib}.  In a Euclidean formulation, the topological model is defined by forcing all plaquettes to be equal to $1$, leaving only Wilson lines tied to an open boundary as physical.  In Hamiltonian formulation, we forbid the usual electric field term in the Hamiltonian and allow only functions of Wilson loops. 

The boundary of $S(l)$ is a union of $d - 2$ dimensional hypercubes and we can define a $Z_2$ group associated with each of these.   $V(S(l))$ transforms under the higher form symmetry by being multiplied with all of the $Z_2$ transformations on its boundary.  Products of $V(S(l))$ variables over closed $d - 1$ dimensional surfaces are invariant under these gauge transformations.   The higher form gauge transformation on a particular $d - 2$ dimensional face of the $d - 1$ dimensional open hypercube $S(l)$ is implemented by the Wilson loop operator $U(\Gamma)$ that includes the link $l$ and encircles that face (see Figure 2 for the $3$ dimensional case).  
\begin{figure}[h]
\begin{center}
\includegraphics[width=01\linewidth]{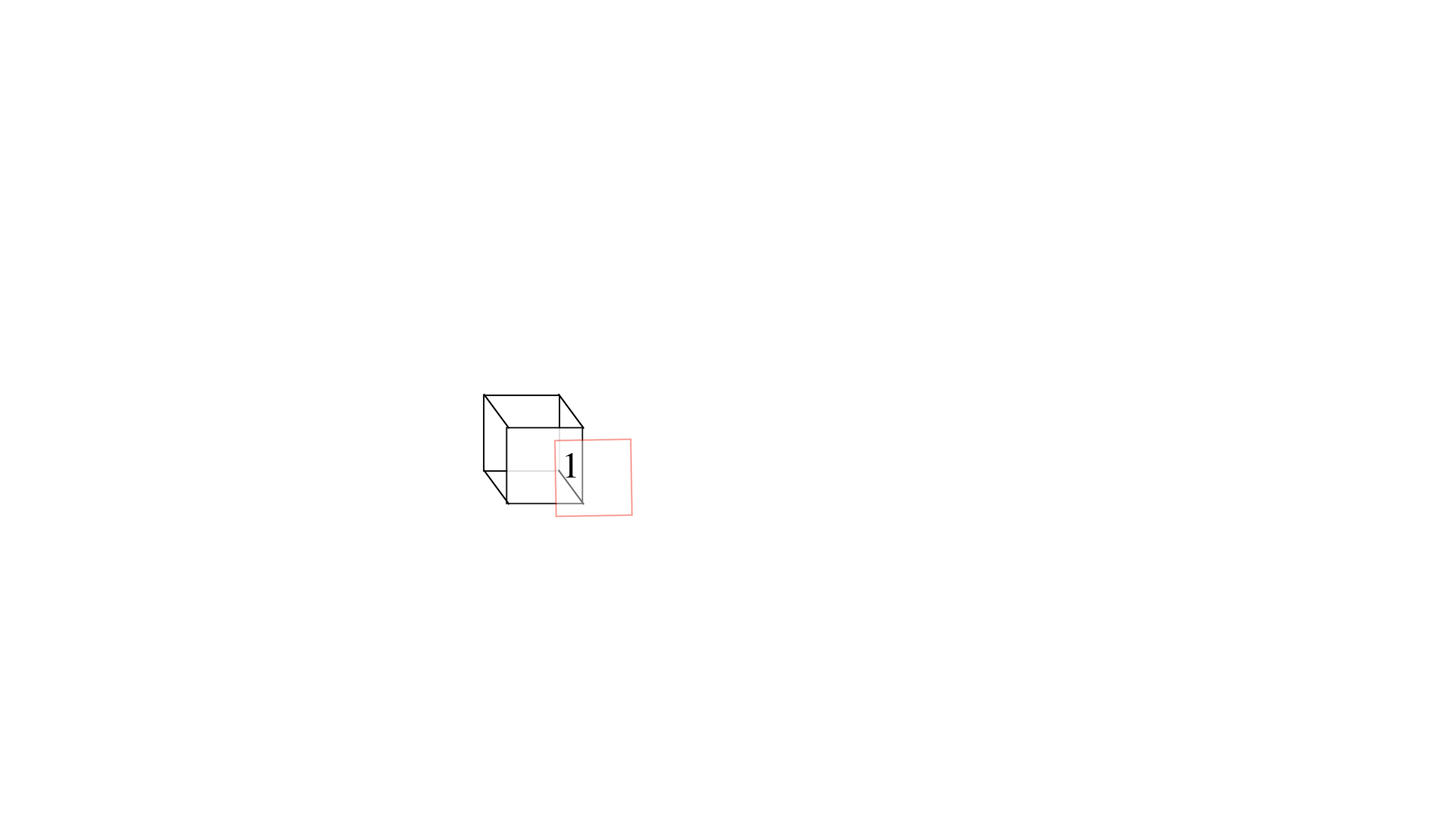}

\caption{The red Wilson loop operator generates the gauge transformation on the encircled link of the indicated electric hypercubic face.} 
\label{fig:fermipaulifigure2}
\end{center}
\end{figure}
The usual electric flux term in the $Z_2$ lattice gauge theory is just the sum of the $V(S(l))$ variables over all $l$ (or what is the same, over all "beginning" lattice sites $m$ in the $(m,n)$ parametrization of links)\footnote{Note that for higher $Z_N$ groups this would be replaced by a sum over $V(S(l)) + V^{\dagger} (S(l))$ . }.  It is not invariant under the higher form gauge symmetry, and therefore forbidden in our construction.   The only gauge invariant operators are $Z_2$ Wilson loops constructed from the $U$ variables, and the Gauss law operators at each lattice site $m$, which are the products of the $V(S(l))$ over all dual hypercubes adjacent to that site.  In a moment we will introduce anti-commuting variables and a Gauss law constraint, which enables us to write all of these gauge theory operators as monomials in fermions.

Now let us introduce dumbbell fields, $\gamma (m,M)$ connecting a point $m$ on the lattice to each of the points on the dual hypercube surrounding it.  These are independent Clifford variables
\begin{equation} [\gamma (m,M), \gamma (n,N)]_+ = 2\delta_{mn} \delta_{MN} . \end{equation}  We impose periodic boundary conditions on the toroidal lattice. 
For each point $M$ on the dual lattice and each lattice direction $\hat{e}$, we have gauge invariant bilinears 
\begin{equation} S(m,m+\hat{e}) = \gamma (m,M) \gamma (m + \hat{e},M) U(m,m + \hat{e}) . \end{equation}  Since these commute with all Wilson loops, they are also invariant under the higher form gauge symmetry.  
We also impose the gauge constraint on the Hilbert space that the product of all the fermionic generators in a dual hypercube is equal to the product of the $V_{S(l)}$ operators on the faces of that hypercube.  Note that we have to choose an order for the $\gamma $ operators in imposing this constraint.  

In order to do this, we first choose an orientation on the dual lattice.  This labels the points at the vertices of a hypercube as $1 \ldots d$ .  We first multiply together $\gamma (m,1) \ldots \gamma (m,d)$.  Now we are at the vertex diagonally opposite $1$ on the hypercube.   As shown in Figure 3 for $d = 2,3$, there is now a unique path through the other vertices, passing through each of them once.  
\begin{figure}[h]
\begin{center}
\includegraphics[width=01\linewidth]{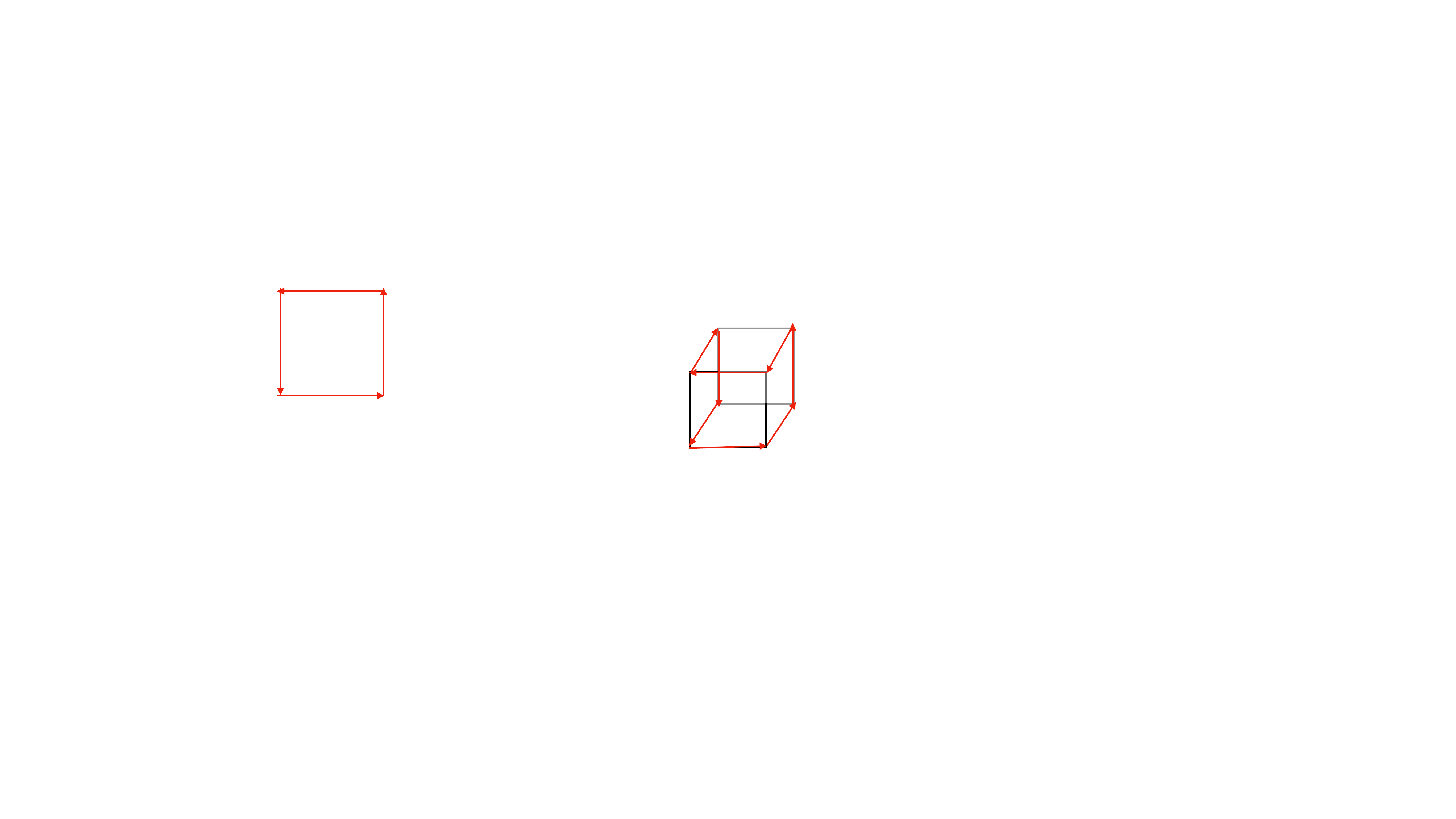}

\caption{The path orderings that define the Gauss law constraint in 2 and 3 dimensions.} 
\label{fig:fermipaulifigure3}
\end{center}
\end{figure}

The resulting product of $2^d$ Clifford elements is called $\Gamma$.
$(1 \pm \Gamma)/2$ defines chiral projectors on the $SO(2^d)$ spinor representation, decomposing it into representations of $SU(2^{d-1}) \times U(1) / Z_{2^{d-1}} $.   The gauge generator is $(- 1)^{N_F}$, with $N_F$ the $U(1)$ generator, so if we introduce complex linear combinations of $\gamma(j,J)$ variables to create and annihilate fermionic particles, the topological gauge group is gauging $(- 1)^{N_F}$.    

The huge symmetry group generated by the Clifford elements at a point has an $SO(d)$ subgroup corresponding to rotations around the point $m$ on the lattice surrounded by the dual hypercube to which the operators $\gamma (m,M)$ belong.  This is not of course a symmetry of the lattice itself, and only a cubic subgroup of it is realized, but we can think of the operators as transforming under it.  They transform as a direct sum of irreducible representations, corresponding to $p$ forms of all ranks between $0$ and $d$.  We have one Clifford variable for each independent component of each $p$ form.  The product of all the Clifford elements is invariant under $SO(d)$ rotations, but can be either even or odd under reflection through the lattice point $m$.  We choose the ordering that is reflection invariant.   

To see these transformation properties, start with $d = 2$.  The sum over all points is the $0$ form and the sum over all points with negative signs for those shifted by one unit in (say) the $\hat{1}$ direction is the two form.  The two differences between points in the $\hat{1}$ and $\hat{2}$ directions are the two components of the $1$ form.  Now take $d = 3$. We can define the $0$ and $3$ forms in a manner completely analogous to the two dimensional case, in terms of their transformation under rotations and reflections.  The one form components are again differences of points in the three lattice directions. The two form components are then defined for each two plane by the rules for $d=2$.   It's now clear how to proceed to general $d$ by induction.

Note that under the $SO(d)$ subgroup of $SU(2^{d-1})$, which we might wish to interpret as performing internal spin rotations on particles localized at points on the lattice, all fields transform in tensor representations.  The model does not have any spinor excitations. This is related to the fact that in the continuum limit, the dual lattice rotation symmetry is really a mixture of a continuum rotation and an emanant internal symmetry\cite{susskind}\cite{seibergtobe}.  Our dumbbell fermions are related to Susskind's staggered fermions and, in the continuum to the Kahler-Dirac equation.

If we multiply the gauge invariant link bilinears around a plaquette, then the Clifford variables all disappear, and we just get the plaquette variables of the $Z_2$ gauge theory.  Thus, the gauge invariant subspace of pure gauge theory is entirely spanned by $S$ operators acting on the function $1$ in the Hilbert space of $L^2$ functions of the Wilson loops of the pure gauge theory.  

We can write other gauge invariant operators in the fermion Hilbert space by constructing bilinears at a single lattice site.  These are the $SO(2^d)$ generators referred to previously. The Hilbert space at a point is $2^{2^d}$ dimensional and can be viewed as constructed from $2^d$ independent q-bits in the usual tensor product construction of Clifford algebras from Pauli matrices.  Link local bilinears in the $S(i,i+\hat{e})$ operators can be rewritten in terms of these Pauli variables.   

The $S(i,i+\hat{e})$ variables do not have a closed algebra.  Rather, they close on an algebra of $Z_2$ strings.   These include Wilson loops of the gauge theory and open strings with fermion fields on their ends.  As noted, fermion fields have a $2^d$ valued index on them, labeling the dual point to which the dumbbell is attached.  It transforms as a vector under the dual $SO(2^d)$ group generated by the operators at a single lattice point.  These fields do not create particle like excitations because they disappear, leaving behind a closed string if they act twice at the same point.  Complex linear combinations behave like $d$ pairs of fermion creation and annihilation operators and can support a $U(2^{d-1})$ symmetry including a particle number $N_F$.  $(- 1)^{N_F}$ is the $Z_2$electric flux through the dual hypercube.  

In fact, all operators in the model can be generated by starting with the link variable $\gamma (m,M) U(m, m + \hat{e}) \gamma (m,M + \hat{e}) $ on a single link, and multiplying it by local operators $ \gamma (n,M)\gamma(n,K)$ from all possible points.   The entire physical operator algebra of the model differs by a single operator from that of an $SO(2^d)$ spin model with a local Hilbert space in the spinor representation.  

So far we have not made a commitment to any particular form of the Hamiltonian.  The system we've defined has an additional $Z_2$  $ d - 1$ form gauge symmetry, whose gauge potentials are the $V_{S(l)}$ variables defined on faces of dual $d - 1$ cubes.  Gauge invariant functions of these variables are products of $V$'s over closed $d - 1$ surfaces.  These all detect open string operators, but only topologically, if the two ends of the string are on opposite sides of the surface.   Thus, if we impose $d - 1$ form gauge invariance on the Hamiltonian the fermionic strings are purely topological in nature.  As noted above, the usual electric field strength term in $Z_2$ lattice gauge theory would violate the $d - 1$ form gauge invariance.  

It is important to add a term that depends on the $Z_2$ Wilson loops to the Hamiltonian.  Of course, since Clifford variables square to one, a Wilson loop is equivalent to a fermion bilinear $\gamma (m,M) U_W (m,m) \gamma (m,M)$. A general quadratic fermion Hamiltonian involving terms that are maximally local on the lattice takes the form.
\begin{equation} \sum_{M,m,\hat{e}}  \gamma (M,m) [\gamma (M, m + \hat{e}) U(m,m + \hat{e}) + c \gamma (M + \hat{e}, m)]  + h \sum_{plaquettes\ P} U(P) . \end{equation}  We've also imposed invariance under cubic rotations and reflections in both the lattice and the dual lattice. 
The ground state energy of the fermions has an expansion in Wilson loops of the gauge theory, which are all products of single plaquette loops.  It can be solved exactly when the plaquettes are uniform, so a large $h$ expansion leads to a systematic perturbation theory in terms of free fermions.   

An extremely important identity is
\begin{equation} \gamma (m,M)\gamma(m,K) \gamma(m+\hat{e},M)\gamma(m+\hat{e},K) = \gamma (m,M)\gamma(m+\hat{e},K) \gamma(m,K) \gamma(m+\hat{e},M) . \end{equation}  This means that a local interaction between purely bosonic spins can be viewed as separating two pairs of fermion fields.  As noted above, by iterating the spin interaction we can create arbitrarily long topological fermionic strings.  To put this in other terms, we could have formulated models without the explicit fermion kinetic term linear in the $S$ operators, as models of $d$ Pauli spins interacting locally on the lattice $m$ with no mention of fermions or fluxes, but their underlying algebra has a hidden fermionic structure with a topological $Z_2$ gauge invariance.  

It is clear that the dumbbell fermions we have constructed are, from the point of view of the dual lattice, a real version of the staggered fermions of Susskind\cite{susskind} which are in turn related to the Kahler-Dirac equation\cite{kahler}\footnote{I've known about the connection between Susskind fermions and Kahler fermions since\cite{dothan}, as well as the connection to the topological twist of SUSY gauge theories\cite{witten} that makes them into topological field theories.  S. Catterall\cite{catterall} has written extensively on this subject and has recently made the fascinating point that a gravitational anomaly of Kahler fermions can be seen on the lattice\cite{catgravanom}.  There's also likely to be a connection to the work of\cite{moorebelov}.}.   This implies that when we take the continuum limit of appropriate Hamiltonians, the geometrical lattice symmetries are revealed to be discrete combinations of continuum symmetries under which the fermion fields transform as spinors, and a continuous internal symmetry acting on a multiplet of spinor fields. In most applications, Susskind fermions are endowed with extra indices and the resulting symmetries coupled to gauge fields.  One may anticipate an interesting interplay between the bosonization proposed here and those applications.

\vskip.3in
\begin{center}
{\bf Acknowledgments }
\end{center}
 The work of T.B. was supported by the Department of Energy under grant DE-SC0010008. Rutgers Project 833012.  Conversations with N. Seiberg about $Z_2$ gauge theory and Susskind fermions are gratefully acknowledged.




\end{document}